\documentstyle[12pt,aaspp4]{article}
\begin{document}

\title{The Gravitational Lens Candidate FBQ 1633+3134\altaffilmark{1}}

\author{
 Nicholas D. Morgan\altaffilmark{2},
 Robert H. Becker\altaffilmark{3,4},
 Michael D. Gregg\altaffilmark{3,4},
 Paul L. Schechter\altaffilmark{2},
 Richard L. White\altaffilmark{5}
}

\altaffiltext{1}{Based on observations carried out in part at the MDM
 Observatory, the W. M. Keck Observatory and the National Radio Astronomy 
 Observatory (NRAO) Very Large Array (VLA).  The NRAO is a facility of the 
 National Science Foundation operated under cooperative agreement by 
 Associated Universities, Inc.}

\altaffilmark{2}{Department of Physics, Massachusetts Institute of
 Technology, Cambridge MA 02139; ndmorgan@mit.edu, schech@achernar.mit.edu}

\altaffilmark{3}{Physics Department, University of California, Davis, CA 95616}

\altaffilmark{4}{Institute of Geophysics and Planetary Physics, Lawrence Livermore National Laboratory, PO Box 808, L-413, Livermore, CA 94551-9900; bob@igpp.ucllnl.org, gregg@igpp.ucllnl.org}

\altaffilmark{5}{Space Telescope Science Institute, 3700 San Martin Drive, Baltimore, MD 21218; rlw@stsci.edu}

\begin{abstract}

We present our ground-based optical imaging, spectral analysis, and 
high resolution radio mapping of the gravitational lens candidate
FBQ 1633+3134.  This $z=1.52$, $B=17.7$ quasar appears double on CCD images
with an image separation of $0\farcs66$ and a 
flux ratio of $\sim3:1$ across $BVRI$ filters.  A single $0.27$ mJy radio 
source is detected at 8.46 GHz, coincident to within an arcsecond of both 
optical components, but
no companion at radio wavelengths is detected down to a 
flux level of $0.1$ mJy ($3\sigma$).  Spectral observations reveal a 
rich metal-line absorption system consisting of a strong \ion{Mg}{2} doublet
and associated \ion{Fe}{1} and \ion{Fe}{2} absorption features, all at an 
intervening redshift of $z=0.684$, suggestive of a lensing galaxy.  
Point spread function subtraction however shows no obvious signs of 
a third object between the two quasar images, and places a detection limit of 
$I \gtrsim 23.0$ if such an object exists.  Although the possibility that FBQ 
1633+3134 is a binary quasar cannot be ruled out, the evidence
is consistent with it being a single quasar lensed by a faint, metal-rich galaxy.

\end{abstract}


\section{INTRODUCTION}

The number of quasars multiply imaged by gravitational
lenses has steadily increased in the last decade.  There are now 
over 40 well-established cases of galaxy-size gravitational lenses (Kochanek 
et al. 1998).  The usefulness of gravitational
lenses are well known (Kochanek \& Hewitt 1996).  For example, measurements 
of a time-delay between the light curves of individual images can lead to a 
determination of the Hubble constant (Refsdal 1964), and the observed 
frequency of strong lensing events can provide statistical constraints on 
the cosmological constant (Kochanek 1996).  Also, since the observed image
configuration is sensitive to the gravitational potential 
of the lens, gravitational lensing offers a means to probe the matter 
distribution of the lensing object (e.g., Maller et al. 2000).  This provides 
a method to sample the dark matter halos of intermediate redshift galaxies 
which is independent of the galaxy's luminosity.

The identification of a close separation (i.e., $<10\arcsec$) double quasar, 
however, is not confirmation of a gravitational lens.  An alternative
explanation is a physical binary quasar (e.g., Kochanek et al. 1999, 
Mortlock et al. 1999).  
In principle, the distinction between these two possibilities is clear:
either there is direct evidence for an intervening galaxy between the
quasar images, confirming the lensing scenario, or the spectral 
properties of the pair differ to the extent that the most likely explanation
is that of two separate quasars.  In the latter case, it is usually the 
failure to detect one of the quasars at radio wavelengths that eliminates the
lensing hypothesis (e.g., Djorgovski et al. 1987; Mu\~noz et al. 1998).  
In practice, there are a number of quasar pairs that do not fall 
neatly into 
either category (Kochanek et al. 1999).  The astrometric and spectral 
characteristics of these pairs are typical of gravitational lenses, yet
no direct evidence for a lensing galaxy is found.  The double quasar 
FIRST J163349.0+313412 (hereafter, FBQ 1633+3134), which we analyze here, is best classified as 
one of these ``border-line'' systems.  In this paper, we present our 
ground-based optical imaging, spectral observations, and radio mapping of 
FBQ 1633+3134, and discuss the likelihood that the system is a 
gravitational lens.

FBQ 1633+3134 (16$^{\scriptsize{\mbox{h}}}$ 33$^{\scriptsize\mbox{m}}$ 
48$\fs$97, +31$\arcdeg$ 34$\arcmin$ 11$\farcs$8; J2000.0) was originally identified as a 
$z=1.52$, $B=17.3$ quasar from the FIRST Bright Quasar Survey (FBQS; Gregg 
et al. 1996).  The target was first detected as a 2 mJy radio source at 20 cm 
as part of the FIRST radio survey (Becker, White, \& Helfand 1995), and 
follow-up spectral observations taken with the Lick 3 m telescope confirmed 
the radio source as a high-redshift quasar (White et al. 2000).  The FBQS
offers a major advantage over fainter quasar surveys for the purpose of 
discovering lensed quasars.  Since the majority of the FBQS targets are
bright ($B<18$), the probability of finding a quasar multiply imaged
by gravitational lensing is enhanced over a sampling of fainter quasars 
due to the magnification bias effect (Turner, Ostriker, \& Gott 1984).  
Using the 2.4 m Hiltner Telescope at the MDM Observatory, we have exploited 
this bias by obtaining optical images of the brightest ($B<18$) and 
more distant ($z>1$) of the FBQS quasars to search for arcsecond scale 
multiple imaging indicative of strong gravitational lensing.  One such 
observing run in 1996 December discovered the quasar FBQ 0951+2635 as a 
sub-arcsecond separation gravitational lens (Schechter et al. 1998).  A 
subsequent observing run was carried out during 1997 June and targeted an 
additional $100$ FBQS quasars in a further search for lenses.  
During this run, a $3$ minute $R$-band exposure of 
FBQ 1633+3134 showed two stellar-like point 
sources separated by $\sim 0\farcs7$, and immediate follow-up exposures 
showed the same separation and similar flux ratios in $BVI$ bandpasses as 
well, providing the initial evidence that FBQ 1633+3134 might be
gravitationally lensed.

We describe our initial optical 
observations and analysis of FBQ 1633+3134 in \S2, as well as 
higher resolution follow-up optical imaging taken immediately afterwards 
using the 3.5 m WIYN telescope.  In \S3, we describe the original FBQS 
spectrum of FBQ 1633+3134, as well as a follow-up spectrum
taken with the Keck 10 m telescope.  While radio observations 
were originally obtained as part of the FIRST survey, we have since obtained 
higher resolution VLA radio imaging of FBQ 1633+3134 following the optical 
detection of the object's double nature.  We present these radio observations 
and their implications in \S4.  Finally, we discuss lens modeling and 
interpretation of the system in \S 5, and \S 6 summarizes our findings 
and conclusions for FBQ 1633+3134.

\section{OPTICAL OBSERVATIONS AND REDUCTION}

\subsection{Initial Optical Imaging}

Initial optical images of FBQ 1633+3134 were taken by one of
us (P.L.S.) with the MDM 2.4 m Hiltner Telescope as part of the optical 
follow-up program for FBQS quasars described above.  Imaging was 
conducted on the nights of 1997 June 1, 2, and 4.  The Tek 1024 $\times$ 
1024 CCD detector (``Charlotte'') was used, with a gain of 3.16 $e^-$ 
ADU$^{-1}$ and a readnoise of 5.45 $e^-$.  The telescope was operated at an $f$
ratio of $f/7.5$, yielding a plate scale of 0$\farcs$2750 per pixel.
Sky conditions were marred by the presence of intermittent clouds on the 
nights of the 2$^{nd}$ and 4$^{th}$, although observations reported 
here were taken during photometric periods.  Multiple $3$ minute 
exposures of FBQ 1633+3134 were taken using the 
Schombert $BVRI$ filter set, with seeing conditions that ranged from 
0$\farcs$8 to 1$\farcs$0 FWHM.  Multiple $BVRI$ observations of the Landolt 
standard field PG 1633 
(Landolt 1992) were also taken on the night of the 2$^{nd}$.  
A summary of the MDM observations for FBQ 1633+3134 
are presented in Table 1.  Figure 1 shows a $5\arcmin$ square field of the 
target quasar and nearby stars from one of the $I$-band frames.

All CCD frames were bias-subtracted, flattened, and trimmed using the 
VISTA reduction program.  The flatfield frames consisted of twilight sky 
exposures taken on June 1, and were removed of cosmic rays using AUTOCLEAN, 
a program written and kindly provided by J. Tonry.  The initial $3$ minute 
$R$-band exposure of FBQ 1633+3134 showed it to be noticeably misshapen as 
compared to other stars in the field, with an ellipticity of $\sim 0.6$ 
(Fig. 2).  This large ellipticity is consistent with a comparatively bright 
companion at less than an arcsecond away from the brighter optical component.  
There is also a faint point source $\sim 3\arcsec$ to the SW of the elongated 
image, with a peak flux of $\sim 1\%$ of the peak flux of the target quasar.

Preliminary photometry for the MDM data was performed using DoPHOT
(Schechter, Mateo, \& Saha 1993), and was able to split FBQ 1633+3134 as 
a double image in all bandpasses.  We fit the elongated image with 
two empirical point spread functions (PSFs), using a variant of the DoPHOT 
program designed to deal with close, point-like and extended objects 
(Schechter \& Moore 1993).  Hereafter, we will refer to the brighter and 
fainter components of the two PSF fit as components A and B, respectively.  We 
simultaneously fit for the presence of the faint SW object
as well (hereafter, component C), to ensure that this object did not 
affect the photometric solutions for the main components.  Star \#5 in Fig. 1 
served as the empirical PSF.

For our photometric fits, we have simultaneously solved 
for the relative 
fluxes of the system components and the overall position of the system, but 
have held the relative separations of the components fixed at values
determined from the Hubble Space Telescope (HST) NICMOS imaging of 
E. Falco (PI), obtained as part of the CASTLES program 
(Kochanek et al. 1998) in 1998 August\footnote{The HST/NICMOS images of FBQ 
1633+3134 may be viewed online at the CASTLES homepage at 
http://cfa-www.harvard.edu/castles}.  Using the archival HST/NICMOS images of FBQ 1633+3134, centroid positions for components 
A, B, and C were determined by the authors using gaussian fits to the peaks 
of the flux profiles; this gave a separation between the brighter two 
components of $0\farcs663$.
Appropriate steps were then taken to account for the 
relative orientations and plate scales of the MDM and HST/NICMOS detectors.  
We report our photometric solutions for the MDM data using fixed relative 
positions for the three components in Table 2.  Error bars in 
Table 2 are statistical ($1\sigma$) errors obtained from the frame-to-frame 
scatter for each filter.

The broadband magnitude differences between components A 
and B show little variation across wavelength.  The photometric 
solutions reported in Table 2 correspond to A:B flux ratios of 2.45 in $B$, 
and 3.25, 3.24, and 3.36 in $V, R,$ and $I$, respectively.  This is consistent
with components A and B having similar spectral energy distributions (SEDs) at 
optical wavelengths, as would be expected if the system consisted of two 
quasars (Mortlock et al. 1999), or a single quasar multiply imaged by 
gravitational lensing.  
Component B has a $B-V$ color index of $0.114$, and $V-R$ and $V-I$
color indices of $0.356$ and $0.691$, with uncertainties of $\sim 0.04$ for 
each index.  The $B-V$ index is consistent with an A star, while 
the $V-R$ and $V-I$ indices are consistent with a late F to early G star.
The color of component B is therefore inconsistent with it being a
foreground Galactic star.  The $B-V$, $V-R$, and $V-I$ color indices for
component A are $0.422$, $0.353$, and $0.725$, with uncertainties of $\sim 0.02$
for each index.  Although these colors are consistent with a late F star,
spectral observations presented in \S 3 make it highly unlikely that the brighter
component is a Galactic star.

The photometric solutions reported in Table 2 also shed light on
the nature of component C.  Component C has $B-V$, $V-R$, and $V-I$ color 
indices of $0.95, 0.50,$ and $0.95$ respectively, with ($1\sigma$) statistical 
errors of $\sim 0.2$ for $B-V$, and $\sim 0.1$ for $V-R$ and $V-I$.
These colors are consistent with a late G to early K star.  
In addition, if we suppose components A and B do indeed arise from 
gravitational lensing, then the relative position of component C with respect 
to components A and B is 
inconsistent with it being either a third image of the lensed quasar or the 
lensing galaxy.  Because it also it appears as a point 
source in our optical imaging, we conclude that
component C is a foreground 
Galactic star and is not physically associated with the two brighter 
components of the system.

\subsection{Astrometry and Photometric Calibration}

The apparent magnitudes for FBQ 1633+3134 reported in Table 2 
were calibrated using offsets
and color terms obtained from observations of the Landolt standard field 
PG 1633 (Landolt 1992), where we have used extinction coefficients taken from 
the Kitt Peak direct imaging manual ($k_B = 0.25, k_V = 0.15, k_R = 0.10, 
k_I = 0.07$; Massey et al. 1997).  DoPHOT yields flux ratios for all objects 
in the field with respect 
to the PSF template star.  Once the PSF star has been calibrated onto 
the standard system, apparent magnitudes for the remainder of objects 
in the field are straightforward to obtain.  We have also obtained astrometric
solutions for reference objects in the field relative to the
Automatic Plate Measuring (APM) astrometry of McMahon \& Irwin (1992) using
one of the $I$-band frames.  To aid in the calibration of future observations,
Table 3 presents our photometric and astrometric solutions 
for 8 reference stars within a 5 arcminute radius of the target quasar. 
The positional uncertainties in the astrometric solutions are $0\farcs4$
($1\sigma$).

\subsection{Follow-Up Optical Imaging}

To probe components A and B for possible signs of a 
lensing galaxy, higher resolution follow-up imaging of FBQ 1633+3134 was 
taken by C. Bailyn immediately following the MDM run.  On 1997 June 5, 
a total of 6 CCD exposures of FBQ 1633+3134 were taken (2 each 
in Harris $B$ and $R$, 
and 1 each in $V$ and $I$) using the WIYN 3.5 m telescope at the Kitt Peak 
National Observatory.  The 2SKB 2048 $\times$ 2048 detector was used, 
with a gain setting of 2.8 $e^-$ ADU$^{-1}$, a readnoise of 8 $e^-$, and a 
plate scale of 0$\farcs$1971 pixel$^{-1}$.  Seeing conditions for the series of
exposures were 0$\farcs$7 FWHM, slightly better than for the MDM run.
Each exposure lasted 120 s.  These images were bias-subtracted, 
flattened, and trimmed using standard IRAF procedures, and the $B$- and 
$R$-band images were stacked using integer pixel offsets before obtaining 
photometric solutions.  PSF fitting was then carried out using the identical 
procedure as for the MDM data, with star \# 5 again providing the empirical 
PSF.

Results from PSF analysis, again using fixed relative positions taken from 
the HST/NICMOS data, yield A:B flux ratios of 2.67 in $B$, and 3.20, 3.21, 3.28 in 
$V$, $R$, and $I$.  These results agree well with the corresponding MDM flux 
ratios for the two components.  In Figure 2, we show an excised portion of the 
$I$-band frame, along with residuals after empirical PSF subtraction for 
the $V$ and $I$ filters.  Tick marks for the residual panels indicate the
centroid locations of components A (top-right) and B (bottom-left); the 
orientation of the system is identical in all four panels.  
We do not show the $R$-band residuals, as the
$R$-band PSF was discovered to vary slightly over the face of the detector,
which masked any residual signature from FBQ 1633+3134 that may have been 
present due to additional sources of flux in the system.   The ellipticity
of the $R$-band PSF increased from $\sim 0.01$ to $\sim 0.06$ for stars
sampled from the center to the periphery of the detector, leaving an
FBQ 1633+3134 residual pattern $\sim 1\%$ of component A's brightness.
The $R$-band PSF-induced residual pattern seen for components A and B of 
FBQ 1633+3134 was similar to the patterns observed for isolated stars in 
the field.

Visual inspection of our ground-based $B$-, $V$- and $I$-band residuals show no
obvious signs of unaccounted flux in the system.  Although we did not perform 
a rigorous PSF subtraction of the HST/NICMOS data, a
cursory inspection of the CASTLES image also shows no obvious indication of a
lensing galaxy.  If a lensing galaxy 
were located between components A and B, its presence would be most evident 
in our data in $I$-band, where the quasar-galaxy flux contrast would 
be more equal.
To quantify the $I$-band magnitude limit for our non-detection, we
inserted a series of gaussian flux profiles between components A and B, representative 
of an intervening 
galaxy, and investigated the residual patterns that emerged after
fitting and subtracting two empirical PSFs as described above.  The location 
of the gaussian profile was dictated by the singular isothermal sphere 
model (see \S 5), with a FWHM of $0\farcs71$ (the average seeing for the 
$I$-band exposure).  Again holding the relative separations of components
A and B fixed at the HST/NICMOS values, we were able to detect $5\sigma$
residuals for a third object down to $I=23.0$, where $\sigma$ is the 
rms sky noise of the exposure.  We therefore rule out the presence
of an $I<23$ magnitude point source at the suspected lensing galaxy position.
The significance of this non-detection is discussed in \S 5.

\section{SPECTROSCOPY}

A spectrum of FBQ 1633+3134 had been originally obtained by one of us (R.B.)
using the Kast Double Spectrograph at the Lick Observatory on 1996 June 10 
as part of the FIRST Bright QSO Survey (see White et al. 2000).  The spectrum
left no doubt that FBQ 1633+3134 was a quasar, clearly showing redshifted 
\ion{C}{4}, \ion{C}{3]}, and \ion{Mg}{2} emission 
features.  Several metal-line absorption features were also detected from an 
intervening line-of-sight absorber.  However, the signal to noise (S/N) of 
the Lick spectrum was only $\sim 10$ at blue wavelengths, which hampered 
the absorption analysis.

We have since obtained a higher S/N spectrum using the Echellette Spectrograph
and Imager (ESI; Epps \& Miller 1998) at the Keck Observatory on 
2000 March 9.  The wavelength
range of the observations was from 3700 \AA$\;$ to 8700 \AA, with a dispersion
that ranged from 0.7 to 11 \AA$\;$ from blue to red wavelengths.  The slit
orientation was held at a PA of $40\arcdeg$ E of N, roughly perpendicular 
to the image separation.  The seeing FWHM was 0\farcs8 --  
insufficient to resolve the two components.  The 600 s spectrum is shown in 
Figure 3.  

Prominent absorption and emission 
features for the FBQ 1633+3134 spectrum are presented in Table 4.  The source quasar has a 
redshift of 1.513(2) based on gaussian fits to peaks of the \ion{C}{3]} and 
\ion{C}{4} emission features.  The S/N of the Keck spectrum ranges from $\sim 
50$ to $\sim 90$ from red to blue wavelengths, allowing the 
detection of an intervening metal-line absorption system at $z_{abs}=0.684$.  
The \ion{Mg}{2} $\lambda \lambda$2796, 2803 \AA$\;$doublet is clearly resolved, 
as is a series of \ion{Fe}{2} $\lambda \lambda$2383, 2586, 2600 
\AA$\;$and \ion{Fe}{1} $\lambda$2474 \AA$\;$absorption lines.  The 
\ion{Mg}{2} doublet has a total rest-frame equivalent width of 2.4 \AA.  
A strong (1.7 \AA$\;$equivalent width) absorption feature 
at an observed wavelength of $\lambda$3708.9 \AA$\;$was also detected in the Lick
spectrum, which may correspond to \ion{Ca}{1} $\lambda$2201 \AA$\;$at 
the slightly higher redshift of $z_{abs} = 0.685$.
The detection of a strong \ion{Mg}{2} and Fe absorption system is evidence for 
a damped Lyman $\alpha$ system (DLAS) at the absorber 
redshift.  A DLAS at intermediate redshift is likely
to be the dusty, metal-rich disk of a late-type galaxy (Boisse et al. 1998).

The high S/N of the Keck spectrum provides a means to 
rule out the possibility that either component is a Galactic star.  
In general, the equivalent width of a spectral absorption feature is reduced 
by a factor of $(1+f)^{-1}$ when flooded with continuum flux from a companion 
object, where $f$ represents the ratio of continuum intensities of the 
brighter object to the fainter object exhibiting the absorption feature.  
Prominent stellar absorption features typically have equivalent widths 
that are of order $1$ \AA$\;$(Jacoby et al. 1984).  Thus, if the fainter optical component 
was indeed a Galactic star, the 3:1 optical flux ratio would predict stellar 
absorption features of typically $0.25$ \AA$\;$equivalent width.  The 
equivalent width associated with the rms noise in the Keck spectrum ranges 
from $0.01$ \AA$\;$in the blue to $0.1$ \AA$\;$in the far red.  The Keck 
spectrum therefore has a high enough S/N to confidently detect stellar 
absorption features if they were present. None of the common stellar absorption features
are detected at the indicated 
positions (dashed lines) of Figure 3, which argues against either component 
being a foreground Galactic star.  The simplest explanation for the combined 
spectrum is that we are seeing the light from two identical redshift 
quasars.

\section{RADIO OBSERVATIONS}

FBQ 1633+3134 is a
milli-Jansky radio source at 20 cm in the FIRST survey.  The survey
was carried out using the Very Large Array (VLA) in the B configuration, 
providing a resolution of $\sim 5 \arcsec$ (Becker et al. 1995).   The 
original radio observations lack the required resolution to probe for 
subarcsecond companions to the target quasar.  Once optical observations had 
revealed the quasar's double nature, higher resolution VLA imaging was 
obtained.

On 1998 March 14, follow-up VLA observations were
carried out by one of us (R.B.) at 8.46 GHz while the VLA was in 
the A configuration, providing an angular resolution of $0\farcs2$.  
Integration time was 60 minutes.  Only one source was securely 
detected, with a peak intensity of 0.27 mJy and an rms noise level in 
the field of 0.03 mJy.

If we assume the radio emission comes from the brighter component of the 
optical pair, then the optical flux ratio of $\sim 3:1$ would predict a 
peak radio intensity for the fainter component of $\sim .1$ mJy.  The 
failure to detect a radio component is therefore significant at the $3\sigma$ 
level.  

The position of the radio source from the high resolution VLA imaging
was determined to be 16$^{\scriptsize{\mbox{h}}}$ 33$^{\scriptsize\mbox{m}}$ 
48$\fs$943, +31$\arcdeg$ 34$\arcmin$ 11$\farcs$19 (J2000.0), 
with a positional uncertainty of $\sim 0.2\arcsec$.  This differs from our 
ground-based optical positions of component A (16$^{\scriptsize{\mbox{h}}}$ 33$^{\scriptsize\mbox{m}}$ 
48$\fs$974, +31$\arcdeg$ 34$\arcmin$ 11$\farcs$85) by $\sim$0$\farcs$8 and 
for component B (16$^{\scriptsize{\mbox{h}}}$ 33$^{\scriptsize\mbox{m}}$ 
49$\fs$020, +31$\arcdeg$ 34$\arcmin$ 11$\farcs$44) by $\sim$1$\farcs$0.  The APM
catalog position for the composite AB source shows the same difference.  We 
also queried the U.S. Naval Observatory
A2.0 (USNO-A2.0) catalog, which uses the Precision Measuring Machine (PPM) 
astrometry of Monet et al. (1996), and found a similar 1$\farcs$0 offset with 
respect to the radio position.
We are unable to identify with confidence which of
the two optical components corresponds to the radio detection.  
A similar $1\arcsec$ discrepancy between 
VLA and optical astrometric solutions was also 
identified for the FIRST lensed quasar FBQ 0951+2635 (Schechter et al. 1998).

\section{MODEL AND INTERPRETATION}

The identification of an intervening metal-line absorption system
in the FBQ 1633+3134 spectrum raises the obvious question of whether
this material could serve as the lens.  Although
this possibility is attractive from the lensing point of view, the mere 
presence of an intervening absorption system should not be taken as 
direct evidence of a lensing galaxy.  The observed 
incidence of Mg II absorption systems in quasar spectra is $\sim 1$ 
such system per unit redshift range down to a minimum
equivalent width of 0.3 \AA$\;$(Steidel \& Sargent 1992), so it is 
common to find an intervening metal-line absorber in the spectra of 
high-redshift quasars.  In this section, we explore the likelihood that
the $z_{abs} = 0.684$ absorber could indeed indicate the presence of the 
lensing galaxy, and investigate the properties of the putative lensing galaxy 
as constrained by other observables of the system.  Throughout this section, 
we have adopted an $\Omega_{m}=0.3$, $\Omega_{\Lambda}=0.7$ cosmological model 
and have parameterized the Hubble constant as $H_{o} = 100 h\mbox{ km s}^{-1} 
\mbox{ Mpc}^{-1}$.

We will adopt the singular isothermal sphere (SIS) model to describe the 
gravitational potential of the hypothesized lensing galaxy.  The SIS model is 
described by the line-of-sight velocity dispersion $\sigma$ of the 
lensing galaxy.  The corresponding Einstein radius is 
$\Theta_{E} = 4\pi\left(\frac{\sigma}{c}\right)^2\frac{D_{LS}}{D_{OS}},$
where $D_{LS}$ and $D_{OS}$ are the angular diameter distances from the 
lens to the source and from the observer to the source, respectively (e.g.,
Narayan \& Bartleman 1998).  The SIS produces a two image lens configuration, with an 
angular separation between the two images of $2\Theta_E$.  If the angular
distance between the brighter (fainter) image and the core of the potential
is denoted by $AG$ ($BG$), then the ratio of image distances $AG/BG$ 
is the same as the lensing-induced flux ratio of the two images.  We
further take the $B$-band luminosity of galaxies to be related to their
line-of-sight velocity dispersions according to a Faber-Jackson (1976) 
relationship of the form $L/L_{\star} = (\sigma/\sigma_{\star})^{\gamma}$. 
Following Keeton et al. (1998), we adopt parameters of 
$\sigma_{\star}$ = 220 (144) km/s and $\gamma$ = 4 (2.6) for early-type 
ellipticals (late-type spirals), where $L_*$ corresponds to a $B$-band 
magnitude of $M_{B}^{*} = -19.7 + 5\log h$.

For a given lensing redshift $z_l$, one can use the observed image
separation to estimate the velocity dispersion and luminosity of the
lensing galaxy.  Figure 4 shows the predicted luminosity of the lensing galaxy
as a function of redshift for both an early-type elliptical and a late-type
spiral.  At the redshift of the metal-line absorber (indicated by the dashed 
vertical line in Figure 4), the predicted luminosities for both
galaxy types are comparable to an $L_*$ galaxy; $\sim 0.3 L_*$ for the 
elliptical, and $\sim 1.3 L_*$ for the spiral, with a corresponding 
dark matter velocity dispersion of $\sim 160$ km/s for both galaxy types.  
The observed image 
separation can therefore be reproduced by a slightly overluminous spiral or
underluminous elliptical (with respect to an $L_*$ galaxy) at the absorber 
redshift.  We also use the techniques of Kochanek (1992) to 
estimate the lens redshift probability distribution for the system.  Using 
a critical lens radius of $r = 0\farcs33$ for FBQ 1633+3134, we compute a 
median redshift for an elliptical (spiral) lens galaxy of $z = 0.72$ 
($0.59$), with a $2\sigma$ redshift interval of $0.21 \lesssim z \lesssim 
1.15$ ($0.17 \lesssim z \lesssim 1.04$).  The detected metal-line absorption 
at $z_{abs} = 0.684$ lies within $1\sigma$ of the median lensing 
redshift for either type of galaxy.

Although the redshift and predicted velocity dispersion of any intervening 
galaxy are consistent with gravitational lensing, the predicted luminosity of 
the lens is difficult to reconcile with the detection limits presented in 
\S2.  Using spectral energy distributions for early- and late-type 
galaxies, we predict the apparent magnitude of the lensing galaxy as a 
function of the lensing redshift $z_l$.  To calculate
broadband apparent magnitudes, we adopt SEDs from Lilly (1997, private 
communication), which consist of interpolated and extrapolated curves from 
Coleman, Wu, and Weedman (1980).  No evolution correction was applied to the 
energy distributions.  The SEDs were normalized to their Faber-Jackson $B$-band
luminosities at $4400(1+z_l)$ \AA, which yields the predicted absolute 
magnitudes on the $AB$ magnitude system.  The corresponding apparent magnitudes
are then computed using the cosmological distance modulus relation
\begin{equation}
	m_{AB}(\lambda_{obs}) - M_{AB}(\lambda_{rest}) = 5\log(\frac{D_L}{10 \mbox{pc}}) + 7.5\log(1+z_l),
\end{equation}
where $D_L$ is the angular diameter distance to the lensing galaxy.
The apparent $AB$ magnitudes are finally transformed onto the standard 
$BVRI$ system by subtracting $-0.110, 0.011, 0.199,$ and $0.456$, respectively
(Fukugita, Shimasaku, \& Ichikawa 1995).  The resulting magnitudes are 
accurate to within $\pm$ 1 magnitude.

Figure 5 presents our $I$-band magnitude predictions for both an early-type 
elliptical and late-type spiral galaxy as a function of lensing redshift.
The predicted apparent magnitudes initially become fainter with increasing redshift,
but eventually increase to brighter magnitudes as the lens redshift approaches
the source redshift.  This increase is because the velocity dispersion 
required to produce a fixed image separation increases without bound as the 
lens approaches the source, leading to a corresponding increase in the
lens luminosity.  For FBQ 1633+3134, the $I$-band detection limit of 
$I=23$ from \S2 roughly corresponds to the faintest predicted magnitude
for an early-type elliptical, but $\sim 1.5$ magnitudes fainter than the
faintest prediction for a late-type spiral.  At the absorber redshift of 
$z_{abs} = 0.684$, the predicted magnitudes lie above the detection limit by
$\sim 0.5$ ($\sim 2.0$) magnitudes for an elliptical (spiral) galaxy.
While an early-type elliptical lensing galaxy could conceivable escape our $I$-band
detection limit, a late-type spiral should have been bright enough to detect
in our optical imaging.

\section{SUMMARY AND CONCLUSIONS}

From an observational standpoint, the requirements for a double quasar 
to be confidently classified as a gravitational lens 
can be stated as follows:  First, since gravitational lensing is 
achromatic, the optical flux ratio of the components should (ideally) be
self-consistent across broadband filters.  In practice, the ratios 
will not agree exactly due to differential reddening through the lens galaxy 
(McLeod et al. 1998) and microlensing by stars in the lens galaxy 
(Wisotzki et al. 1993).  More importantly, the radio flux 
ratio, if available, must also agree with 
the optical ratio, again because of the achromatic nature of lensing.  
Second, spectral observations of the quasar pair must support the conclusion 
that both objects are identical redshift quasars with similar (although not 
necessarily identical; Wisotzki et al. 1993) emission and continuum 
properties.  Third, the lensing galaxy itself must be detected, ideally by 
direct imaging,  although weaker arguments based on the presence of 
intervening absorption features in the QSO spectrum have been used in the 
past (e.g., Hewett et al. 1994).  In practice, the first two requirements are 
necessary for a lens classification, while the third (direct imaging of the 
lensing galaxy) is usually sufficient.

FBQ 1633+3134 meets the necessary requirements for a gravitational lens.  
The broadband flux ratio ($\sim 2.5:1$ in $B$, $\sim 3.3:1$ in $VR$, \& $I$) is 
self-consistent across broadband wavelengths.  Although only
one component was detected in radio, the lower limit to the radio
flux ratio of $\gtrsim 3:1$ is not in severe conflict with the 
corresponding optical flux ratio.  Also, the spectral analysis 
argues against either component of the pair being a Galactic star, and the 
unresolved emission profiles from the pair are consistent with two 
identical-redshift quasars.

The sufficient condition that could elevate FBQ 1633+3134 to 
the status of a lensed quasar, direct detection of the lensing galaxy, is not 
realized with our optical imaging.  Subtraction of two PSFs for the
WIYN data presented in \S2 show no significant indication of unaccounted
flux in the system, and place a magnitude limit of $I>23.0$ using
a simple SIS model for the lensing potential.  Although the hypothesized lens
galaxy is not detected directly, a weaker argument for 
its presence can be made from the spectral identification of a rich 
$z_{abs}=0.684$ metal-line absorption system, which is within 
$1\sigma$ of the median lensing 
redshift.  Also, the estimated luminosity of the suspected galaxy
does not differ strongly from an $L_*$ galaxy for either of the galaxy types
considered here.  On the other hand, the predicted apparent magnitude of a 
late-type spiral at the absorber redshift is $\sim2$ magnitudes above the
$I$-band detection threshold,
while an early-type galaxy is only marginally consistent with the detection
limit.

We conclude that FBQ 1633+3134 is a strong candidate for 
a close separation (0\farcs66) gravitational lens.  The evidence presented here suggests 
lensing by a relatively faint ($I>23.0$), $z=0.684$ metal-rich galaxy,
although the binary quasar scenario cannot be positively ruled out.   
Confirmation of the lensing hypothesis can be provided by a deeper radio 
probe for the fainter component of the system.  We have obtained VLA A 
array time for late 2000 with this aim in mind, and intend to present these
results, along with a systematic analysis of the HST data for this system, in 
a future paper.

\acknowledgements

The authors would like to thank Charles Bailyn for obtaining the WIYN 
optical data of FBQ 1633+3134, and Brian Skiff for noticing a V band calibration 
error in an earlier version of this paper.
N.D.M. and P.L.S. gratefully acknowledge 
the support of the U.S. National Science Foundation through grant AST96-16866. 
Some of the data presented here were obtained at the W.M. Keck Observatory, 
which is operated as a scientific partnership among the California 
Institute of Technology, the University of California and
the National Aeronautics and Space Administration.  The Keck
Observatory was made possible by the generous financial support of the
W.M. Keck Foundation.  The FIRST Survey is supported by grants from
the National Science Foundation (grant AST98-02791), NATO, the
National Geographic Society, Sun Microsystems, and Columbia
University.  Part of the work reported here was done at the Institute
of Geophysics and Planetary Physics, under the auspices of the
U.S. Department of Energy by Lawrence Livermore National Laboratory
under contract No.~W-7405-Eng-48.

\clearpage

\figcaption[fig1.ps] 
        {A $3$ minute $I$-band exposure of FBQ 1633+3134 (center crosshairs) 
and surrounding field.  North is up and east is to the left.  The scale of 
the image is shown in the bottom left of the figure.
\label{fig1}}

\figcaption[fig2.ps] 
        {Extracted subrasters of FBQ 1633+3134 taken with the WIYN 3.5 m 
telescope.  North is up and East is to the left.  Each panel is $\sim 
4\arcsec$ wide.  Top left: A $120$ s $I$-band exposure, showing component A 
(NW) and component B (SE).  The small tick mark SW of the pair indicates the 
centroid location of component C (which is too faint to be seen at the given 
contrast level).  Top right:  $I$-band residuals from two empirical PSF 
subtraction of components A and B.  Tick marks indicate the centroid location 
of the two components.  Bottom left: Stacked $B$-band residuals from two 
component PSF subtraction.  Bottom right: $V$-band residuals from two 
component PSF subtraction.  The saturation levels for the residual panels 
are at $\pm 10\sigma$, where $\sigma$ is the respective rms sky noise for 
each image.  Component A has a peak intensity of $\sim580\sigma$, 
$\sim482\sigma$, and $\sim420\sigma$ for the $B$-(stacked), $V$-, and $I$-
band exposures.
\label{fig2}}

\figcaption[fig3.ps] 
        {Composite spectrum of FBQ 1633+3134 A,B taken with the Keck 10 m 
telescope.  Integration time was $600$ s.  Prominent emission features are 
indicated for this $z=1.52$ quasar, as well as the identification of a 
$z_{abs}=0.684$ metal-rich absorption system.  
The lack of any stellar absorption features at the 
indicated wavelengths (dashed lines) argues against either component being a 
foreground Galactic star.
\label{fig3}}

\figcaption[fig4.ps] 
        {Predicted $B$-band luminosity (in units of $L_*$) of the hypothesized 
lensing galaxy as a function of its lensing redshift $z_l$.  Results for a 
late-type spiral and early-type elliptical are shown.  The redshift of the 
intervening absorption system is indicated by the vertical dashed line.  
The corresponding line-of-sight velocity dispersion at the absorber redshift is 
$\sim160$ km s$^{-1}$.
\label{fig4}}

\figcaption[fig5.ps] 
        {Predicted $I$-band apparent magnitudes of the hypothesized lensing 
galaxy as a function of its lensing redshift $z_l$.  Results are 
shown for a late-type spiral (dot-dash line) and early-type elliptical (solid
line).  The $I=23.0$ detection limit for any such galaxy is indicated
by the heavy horizontal line, and the redshift of the intervening absorption
system is shown by the vertical dashed line.
\label{fig5}}





 

\makeatletter
\def\jnl@aj{AJ}
\ifx\revtex@jnl\jnl@aj\let\tablebreak=\nl\fi
\makeatother


\begin{deluxetable}{ccc}
\tablecaption{Summary of MDM Observations (June 1997)
\label{TABLE1}}
\tablenum{1}
\tablewidth{0pt}
\tablehead{
\colhead {Filter} &
\colhead {N$_{img}$} &
\colhead {Average FWHM}}
\startdata
B &  5 & 1$\farcs$00 \nl
V & 12 & 0$\farcs$82 \nl
R &  4 & 0$\farcs$87 \nl
I &  8 & 0$\farcs$87 \nl
\enddata
\end{deluxetable}


\clearpage



 

\makeatletter
\def\jnl@aj{AJ}
\ifx\revtex@jnl\jnl@aj\let\tablebreak=\nl\fi
\makeatother


\begin{deluxetable}{lcccccc}
\tablecaption{MDM Photometric Solutions for FBQ 1633+3134
\label{TABLE2}}
\tablenum{2}
\tablewidth{0pt}
\tablehead{
\colhead {Comp.} &
\colhead {$\Delta$ R.A.} &
\colhead {$\Delta$ Dec.} &
\colhead {B} &
\colhead {V} &
\colhead {R} &
\colhead {I}}
\startdata
A........ & ---              & ---             & 18.046  $\pm$  0.017 & 17.624 $\pm$   0.006 & 17.271 $\pm$   0.013 & 16.899 $\pm$   0.010 \nl
B........ & 0$\farcs$524     & -0$\farcs$406   & 19.019  $\pm$  0.033 & 18.905 $\pm$   0.019 & 18.549 $\pm$   0.037 & 18.214 $\pm$   0.034 \nl
C........ & -1$\farcs$793    & -2$\farcs$282   & 23.268  $\pm$  0.167 & 22.315 $\pm$   0.071 & 21.814 $\pm$   0.037 & 21.370 $\pm$   0.063 \nl
\enddata
\tablecomments{Relative astrometric positions were obtained from the archival 
CASTLES HST/NICMOS imaging.  Magnitude solutions for the 
MDM data were obtained while holding the relative separations fixed at the 
HST values. Error bars are statistical ($1\sigma$) and do not include 
uncertainties in the PSF star calibration.  See note to Table 3.}

\end{deluxetable}


\clearpage



 

\makeatletter
\def\jnl@aj{AJ}
\ifx\revtex@jnl\jnl@aj\let\tablebreak=\nl\fi
\makeatother


\begin{deluxetable}{ccccccc}
\tablecaption{Relative Astrometry and Apparent Magnitudes for Field Stars
\label{TABLE3}}
\tablenum{3}
\tablewidth{0pt}
\tablehead{
\colhead {Object} &
\colhead {$\Delta$$\alpha$ ($^s$)} &
\colhead {$\Delta$$\delta$ ($\arcsec$)} &
\colhead {B} &
\colhead {V} &
\colhead {R} &
\colhead {I}}
\startdata
1 & -3.010 & 115.11 & 18.299 $\pm$ 0.002 & 17.128 $\pm$ 0.001 & 16.407 $\pm$ 0.001 & 15.827 $\pm$ 0.002 \nl
2 & -1.206 &  82.37 & 19.158 $\pm$ 0.002 & 18.537 $\pm$ 0.003 & 18.215 $\pm$ 0.002 & 17.864 $\pm$ 0.003 \nl
3 &  5.580 &  66.18 & 20.557 $\pm$ 0.006 & 19.296 $\pm$ 0.002 & 18.489 $\pm$ 0.002 & 17.775 $\pm$ 0.003 \nl
4 & -1.829 &  63.61 & 19.853 $\pm$ 0.001 & 18.605 $\pm$ 0.002 & 17.776 $\pm$ 0.002 & 17.058 $\pm$ 0.002 \nl
5 &  0.000 &   0.00 & 18.670 $\pm$ 0.000 & 17.187 $\pm$ 0.000 & 16.059 $\pm$ 0.000 & 14.673 $\pm$ 0.000 \nl
6 &  4.275 &  -5.35 & 17.200 $\pm$ 0.002 & 16.225 $\pm$ 0.001 & 15.708 $\pm$ 0.001 & 15.242 $\pm$ 0.002 \nl
7 &  4.216 &  -8.93 & 17.556 $\pm$ 0.003 & 16.417 $\pm$ 0.001 & 15.729 $\pm$ 0.001 & 15.108 $\pm$ 0.002 \nl
8 &  0.885 & -42.38 & 18.702 $\pm$ 0.003 & 17.810 $\pm$ 0.001 & 17.339 $\pm$ 0.002 & 16.904 $\pm$ 0.002 \nl
\enddata
\tablecomments{
Object numbers correspond to the labels shown in Figure 1.
Reported error bars are statistical ($1\sigma$) errors from the 
observed dispersion between frames, and do not include uncertainties in the
calibration of the PSF star (object \#5).  Magnitude uncertainties (1$\sigma$)
for the PSF star are $0.007$, $0.008$, $0.005$, and $0.004$ for $BVRI$ filters,
respectively, and must be added in quadrature to the errors reported above.}
\end{deluxetable}


\clearpage



 

\makeatletter
\def\jnl@aj{AJ}
\ifx\revtex@jnl\jnl@aj\let\tablebreak=\nl\fi
\makeatother


\begin{deluxetable}{ccccc}
\tablecaption{Spectral Analysis for FBQ 1633+3134
\label{TABLE4}}
\tablenum{4}
\tablewidth{0pt}
\tablehead{  
\colhead {Emission} &
\colhead {Absorption} &
\colhead {$\lambda_{obs}$} &
\colhead {$\lambda_{rest}$}&
\colhead {z}
}
\startdata
\ion{C}{4} & ..... & 3889.9 & 1549.3 & 1.5107(5)  \nl
\ion{C}{3]}& ..... & 4799.7 & 1908.7 & 1.5146(11) \nl
\ion{Mg}{2}& ..... & 7064.4 & 2799.5 & 1.5235(5)  \nl
..... & \ion{Ca}{1}& 3708.9 & 2201.4 & 0.6848(4)  \nl
..... & \ion{Fe}{2}& 4012.0 & 2382.8 & 0.6837(2)  \nl
..... & \ion{Fe}{1}& 4165.8 & 2473.9 & 0.6839(3)  \nl
..... & \ion{Fe}{2}& 4354.8 & 2586.7 & 0.6836(2)  \nl
..... & \ion{Fe}{2}& 4377.9 & 2600.2 & 0.6837(1)  \nl
..... & \ion{Mg}{2}& 4707.5 & 2796.4 & 0.6834(3) \nl
..... & \ion{Mg}{2}& 4719.8 & 2803.5 & 0.6835(1) \nl

\enddata
\tablecomments{Numbers in parenthesis are $1\sigma$ error bars.}
\end{deluxetable}

\clearpage

\begin{figure}[h]
\vspace{7.0 truein}
\includegraphics{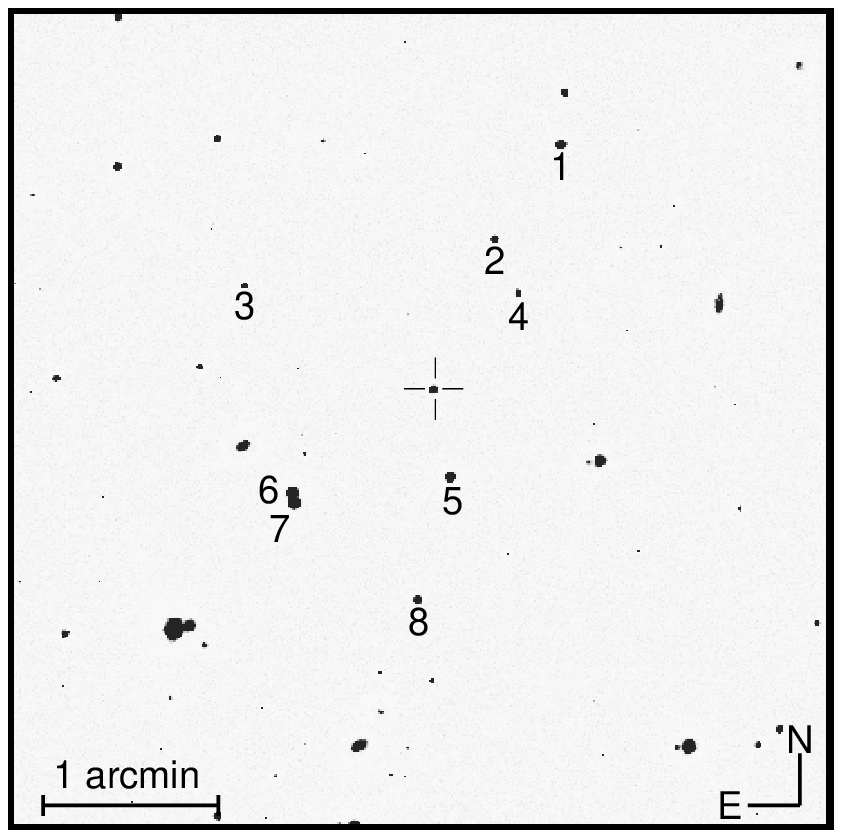}
\end{figure}
\clearpage

\begin{figure}[h]
\vspace{7.0 truein}
\includegraphics{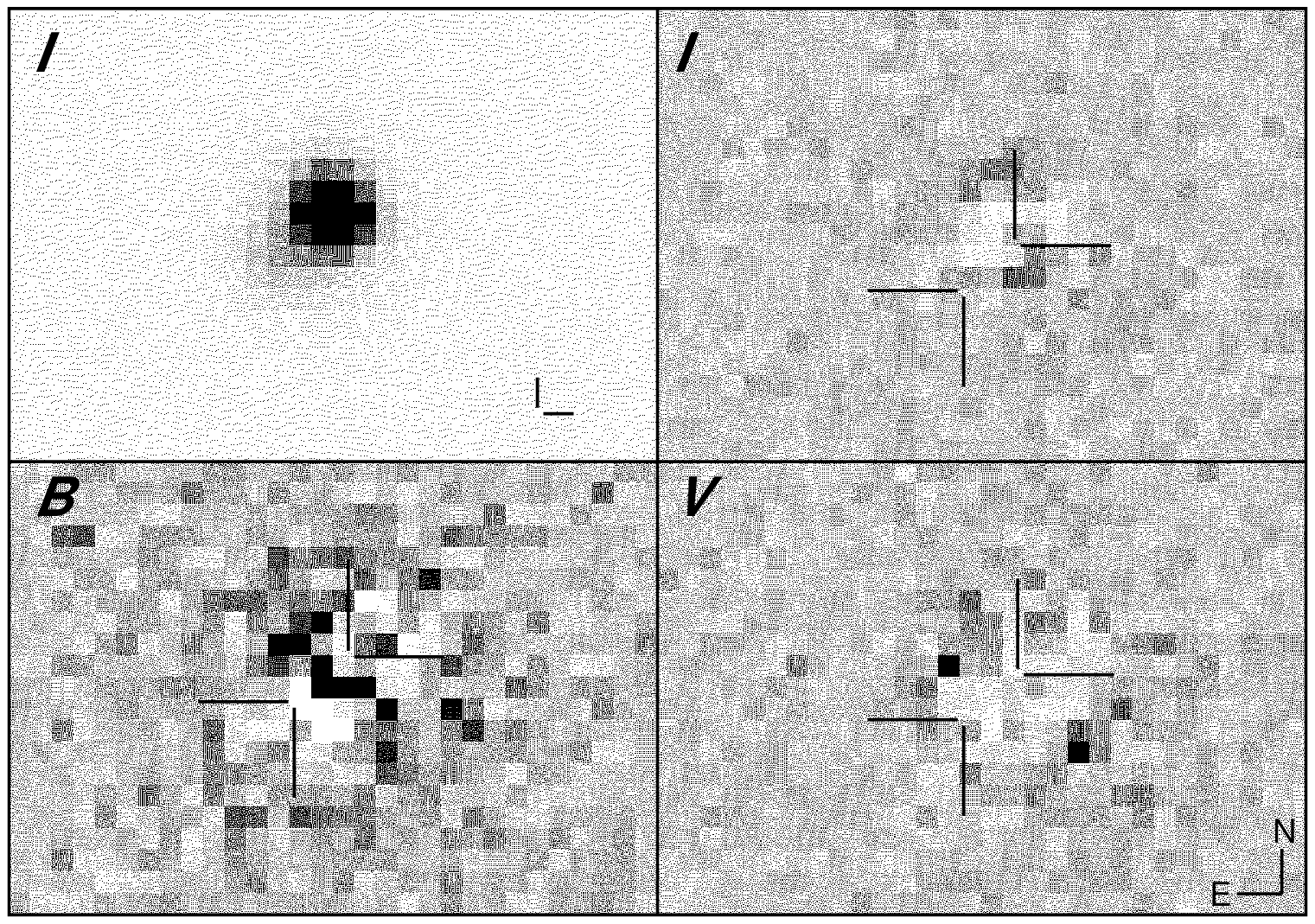}
\end{figure}
\clearpage

\thispagestyle{empty}
\begin{figure}[h]
\vspace{7.0 truein}
\includegraphics{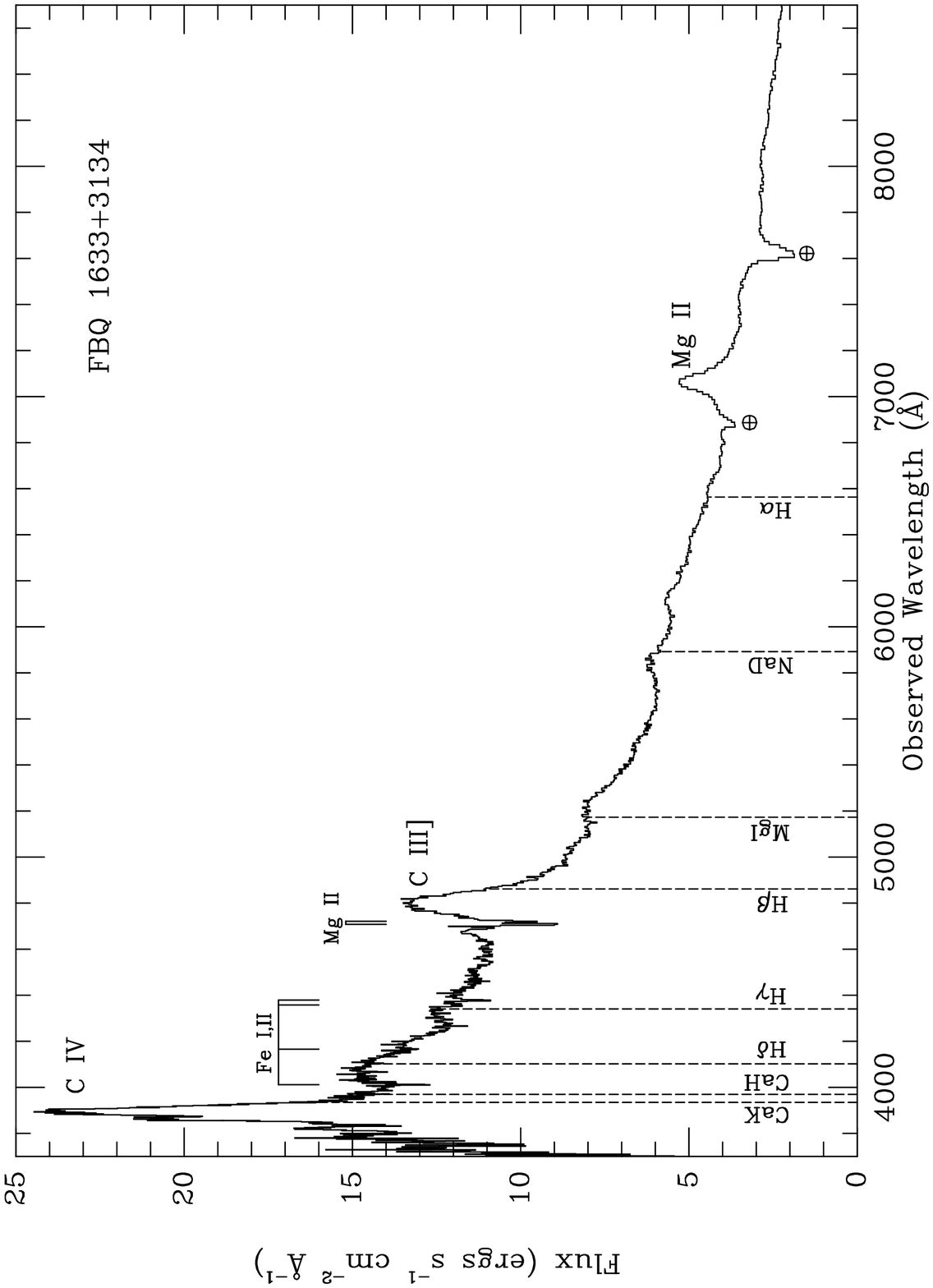}
\end{figure}
\clearpage

\thispagestyle{empty}
\begin{figure}[h]
\vspace{7.0 truein}
\includegraphics{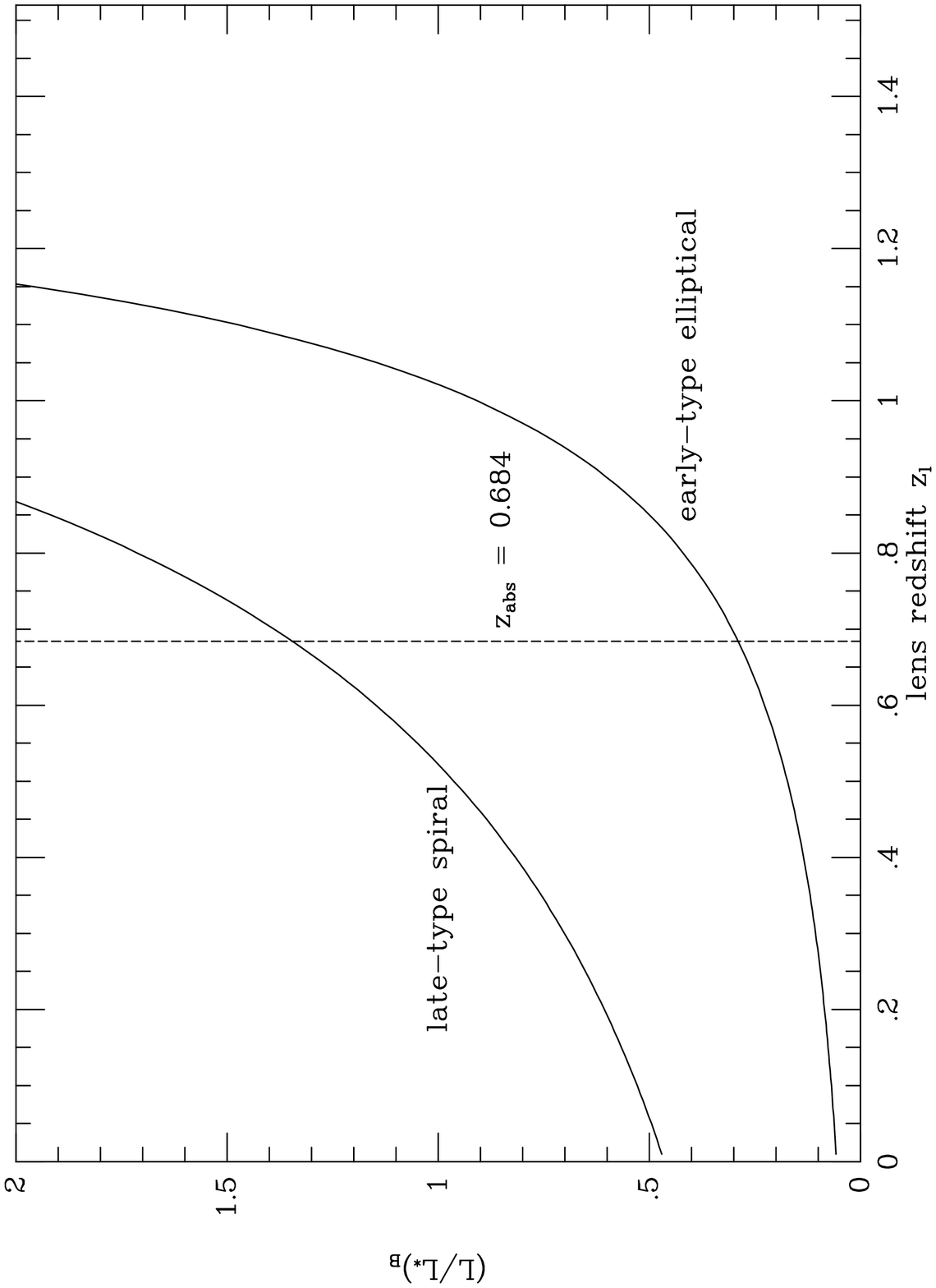}
\end{figure}
\clearpage

\thispagestyle{empty}
\begin{figure}[h]
\vspace{7.0 truein}
\includegraphics{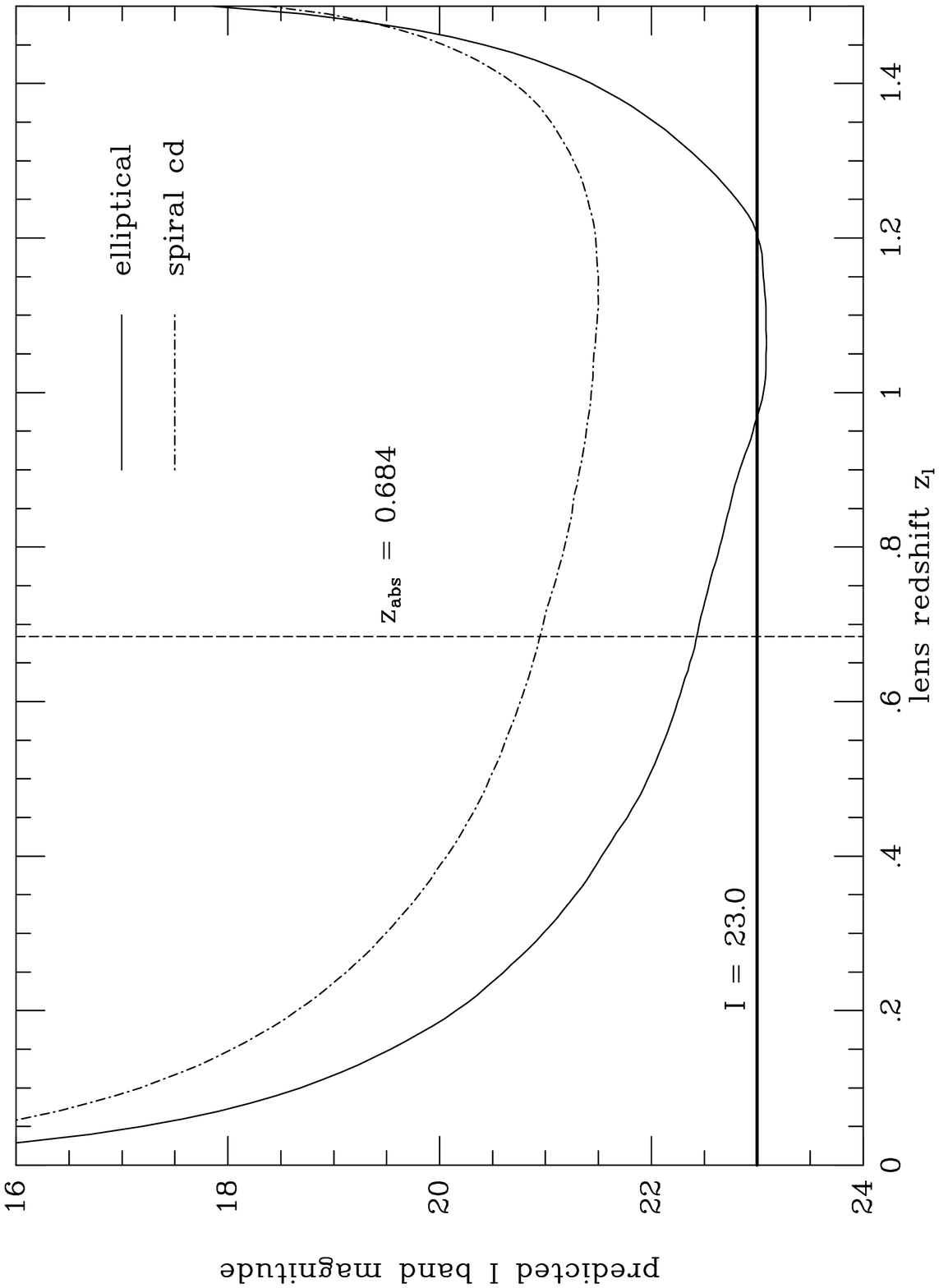}
\end{figure}
\clearpage

\end{document}